# Five Frequently Fatal Freshmen Physics Fantasies

Michael Courtney, Norm Althausen, and Amy Courtney

*Abstract:*
This article describes five common student misconceptions about succeeding in college-level Physics courses: the miracle finish, the soft hearted professor, an extension of high school, weak areas won't be tested, and passing is more important than learning. These fantasies are unique to neither Physics nor freshman, but the challenge of learning Physics brings more certain negative consequences to students retaining these attitudes. Ascribing these student misconceptions to freshmen acknowledges that they diminish with increasing academic maturity.

**Fantasy of the Miracle Finish**
In the movies and television, protagonists are often seen making weak academic or professional efforts (often due to romantic or heroic distractions) and then "buckling down" and working "real hard" for a short time before pulling off the required performance to salvage their grade or job. Experienced Physics educators know it is nearly impossible to learn a full term's worth of Physics in a short time. Unsatisfactory levels of student effort for most of the semester cannot be overcome with improved efforts at the end. One farmer can produce a crop of grain in six months, but six farmers cannot produce a crop of grain in one month.

Miracle finishes in Physics result from the kind of consistent diligence displayed in the hockey movie, "Miracle", that recounts the USA's victory over the USSR's hockey team in the 1980 Olympic Games. The problem solving skills required in most Physics courses represent the formation of abilities in addition to the transmission of information. Acquisition of these skills is more like learning to play a musical instrument than learning many academic subjects. Success requires repeated application of methods demonstrated in class. The coach of USA's 1980 hockey team gave birth to the miracle finish by his exhortations in practice: again, again, again, again …

Students need to know they might have a better chance winning an Olympic gold medal or a Grammy award than passing a Physics course without a significant and steady level of effort from the beginning of the course.

**Fantasy of the Soft Hearted Professor**
Many students finally show up at the professor's door near the end of the semester asking, "What can I do to pass this class?" Often, the only honest answer is, "Work hard from the beginning when you retake the course." This conversation often includes sad excuses for a student's sub-optimal efforts, and compelling reasons why the student "needs" to pass the course this semester.

Students with this fantasy do not believe their grade will be determined simply by applying the policy described in the syllabus to their graded work. Could the positive perception of the professor as a friend deceive students that the grading policy does not apply to them?

Students must internalize that the grading policy will be applied evenly and literally to all students.  The assigned grade will represent learning and efforts as described in the syllabus, rather than the instructor's personal feelings.

**Fantasy that College is a Simple Extension of High School**
Many students begin introductory Physics thinking that high school habits and attitudes will ensure success.  This fantasy of passive learning suggests simply attending class is sufficient and having a book in the lap in front of the television constitutes productive study time.

Students must be whole hearted in actively engaging challenging courses.  Physics is learned by practice more than studying.  The most productive learning time is spent with the pencil moving, with the thoughts fully focused on the classroom discussion, or with the mind and hands fully engaged in the laboratory.  It is folly to think that getting the notes from a friend, the data from a lab partner, or passively reading the material is an adequate substitute for the full first hand experience.

The high school habit of cramming the night before a test is analogous to the fantasy of the miracle finish worked out on a time scale of several weeks rather than the entire term.  Even if there were sufficient time to catch up on several weeks of assignments, productivity would be limited.

**Fantasy that Weak Areas Won't Be Tested**
Some topics and kinds of problems are harder than others.  This engenders avoidance behavior where students concentrate on areas of greater comfort while hoping that weaknesses will not be tested.  Students often maintain this self-deception in spite of explicit exhortations that more challenging material must be learned.

This attitude is subtly expressed in sentiments such as, "Can we have a multiple choice exam?" and, "Will this be on the test?"  Even when told explicitly that they are accountable for material on the exam, students often convince themselves the difficult material will not count for enough points to hurt their grade.

Instructors must effectively communicate the mathematical impossibility of succeeding in a course without competency in certain required areas.  For example, many students do not become proficient in vector analysis and will attempt to pass a course approaching every exam exercise as a one-dimensional problem.  If a course requires competency in this skill, students must learn that problems requiring vector analysis will comprise a substantial portion of exam credit.

**Fantasy that Passing is More Important than Learning**
Some students focus on the college degree that may gain initial entry into a profession rather than the skills and knowledge that bring long-term success. This manifests as the idea that credit in individual courses is more important than the learning it represents.

Instructors must ensure that students understand the training, attitude, skills, and knowledge required in their chosen professions.  We must also carefully consider whether our teaching and evaluation methods reinforce this fantasy or assist students in growing out of it.

The improvement process begins by bringing these fantasies into the light for students.  Fully overcoming these fantasies requires character formation rather than merely the transmission of information.  Old habits need to be replaced with productive attitudes and habits.  This character formation might represent the most valuable long-term learning gained from an introductory Physics course, and we must seek metaphors, illustrations, and methods to empower the required changes.

**About the Authors**
Michael Courtney earned a PhD in Physics from MIT.  He has taught introductory Physics for several years as well as Student Development (College 101).  He is currently on the Physics faculty at Western Carolina University where he serves as director of the Forensic Science program. Chemistry and Physics, Natural Science 230, Cullowhee, NC 28723.  Michael_Courtney@alum.mit.edu

Norm Althausen earned a BA in Geology from the University of California, Davis and an MA in Educational Administration from the University of San Francisco.  He has taught AP Physics and AP Calculus for the last fifteen years at Beachwood High School, Beachwood, OH, as well as being an Adjunct Professor of Physics at Lorain County Community College.

Amy Courtney holds a PhD in the joint Harvard/MIT program in Medical Engineering and Medical Physics.  She has worked as a research scientist at Reebok and the Cleveland Clinic and served on the faculty of the Ohio State University.